\documentstyle [12pt] {article}
\begin{document}
%%%%Start of Text%%%%%%%%%%%%%%%%%%%%%%%%%%%%%%%%%%%%%%%%%%%%%%%%%%%%%%%%%%%%
\pagestyle{empty}
\rightline{\vbox{
\halign{&#\hfil\cr
&NUHEP-TH-93-2\cr
&UCD-93-1\cr
&February 1993\cr}}}
\bigskip
\bigskip
\bigskip
{\Large\bf
	\centerline{$Z^0$ Decay into Charmonium}
	\centerline{via Charm Quark Fragmentation}}
\bigskip
\normalsize

\centerline{Eric Braaten and Kingman Cheung}
\centerline{\sl Department of Physics and Astronomy, Northwestern University,
    Evanston, IL 60208}
\bigskip

\centerline{Tzu Chiang Yuan}
\centerline{\sl Davis Institute for High Energy Physics}
\centerline{\sl Department of Physics, University of California,
    Davis, CA  95616}
\bigskip

\begin{abstract}
In decays of the $Z^0$, the dominant mechanism for the direct
production of charmonium states
is the decay of the $Z^0$ into a charm quark or antiquark
followed by its fragmentation into the charmonium state.
We calculate the fragmentation functions describing the splitting
of charm quarks into S-wave charmonium states to leading order in the
QCD coupling constant.  Leading logarithms of $M_Z/m_c$
are summed up using Altarelli-Parisi evolution equations.
Our analytic result agrees with the complete leading order calculation
of the rate for $Z^0 \rightarrow \psi c {\bar c}$.
We also use our fragmentation functions to calculate
the production rate of heavy quarkonium states in $W^\pm$, top quark,
and Higgs decays.
\end{abstract}

\vfill\eject\pagestyle{plain}\setcounter{page}{1}

{\bf \centerline{Introduction}}

Among the rare decay modes of the $Z^0$ predicted by the Standard Gauge Theory
are ones whose final states include charmonium.
Of particular importance are the $^3S_1$ charmonium
states $J/\psi$ and  $\psi'$,
since their decays into lepton pairs provide easily identifiable
experimental signatures.  The dominant production mechanism for
$\psi$ and $\psi'$ is the decay of $B$ hadrons; in fact, this serves as a
signature for $B$ hadron production in $Z^0$ decay.  The direct
production of $\psi$ and $\psi'$ is therefore important in $Z^0$ decays as a
background to $B$ physics.  It is also
of interest in its own right, since it involves both short distance and
long distance aspects of quantum chromodynamics (QCD).  The production of a
charm quark and antiquark with small relative momentum in $Z^0$ decay is
a short distance process with a characteristic length scale that can range from
$1/M_Z$ to as large as $1/m_c$.  The subsequent formation of a bound state
{}from the $c {\bar c}$ pair is a long distance process involving all the
complications of nonperturbative QCD.  The methods of perturbative
QCD can be used to calculate the production rates provided
that it is possible to systematically separate the short distance
effects from the long distance effects.

Most previous work on charmonium production in $Z^0$ decay
\cite{gkpr,keung,abr}
has focused on {\it short distance} processes
in which the $c {\bar c}$ pair that form the $\psi$ is produced
with a transverse separation of order $1/M_Z$.
 Long distance effects involved
in the formation of the bound state are factored into the nonrelativistic
radial wavefunction at the origin $R(0)$. The best example of a short
distance process is $Z^0 \rightarrow \psi g g$,
which has a branching fraction of about $10^{-7}$.
This small branching fraction can be partly
attributed to a factor of $|R(0)|^2/(m_c M_Z^2)$, which represents the
probability for a $c {\bar c}$ pair that is produced in a region of size
$1/(m_c M_Z^2)$ to form a bound state.  This probability factor
suppresses the branching fractions for short distance processes
by $m_c^2/M_Z^2$,
so that they can be neglected in the limit $M_Z/m_c \rightarrow \infty$.

As pointed out by K\"uhn and Schneider \cite{ks},
the direct production of charmonium in $Z^0$ decay will be dominated
not by short distance processes but by {\it fragmentation} processes.
The fragmentation mechanism
is the decay of the $Z^0$ into a final state that includes a high energy
quark or gluon, followed by the splitting of that parton
into the charmonium state plus other partons.
In the fragmentation mechanism, the $c$ and ${\bar c}$ that form the charmonium
state are produced with a separation of order $1/m_c$.  The probability that
they form a bound state is proportional to $|R(0)|^2/m_c^3$.
The branching ratio for such a process is therefore not
suppressed by the factor $m_c^2/M_Z^2$ associated
with short distance processes.  The  fragmentation of a parton is described by
a fragmentation function $D(z,\mu)$, which gives the probability
for a parton with invariant mass less than $\mu$ to split into the charmonium
state with longitudinal momentum fraction $z$.  It was recently shown that
the fragmentation functions for the splitting of
partons into heavy quarkonium states can be
calculated using perturbative QCD \cite{by}.  The fragmentation functions
$D_{g \rightarrow \psi}(z,\mu)$ and $D_{g \rightarrow \eta_c}(z,\mu)$
that describe the splitting of gluons into S-wave quarkonium states were
calculate to leading order in $\alpha_s$ at the scale $\mu = m_c$.
They were evolved to larger scales $\mu$ by using
Altarelli-Parisi evolution equations, which sum up leading
logarithms of $\mu/m_c$.  The production of $\psi$ in $Z^0$ decay
{}from the splitting of virtual gluons has been considered by
Hagiwara, Martin, and Stirling \cite{hms}, but they did not organize
the calculation in terms of fragmentation functions and were thus unable
to sum up leading logarithms of $M_Z/m_c$.

The production rate of $\psi$ via the process
$Z^0 \rightarrow \psi c {\bar c}$ has been calculated by Barger, Cheung,
and Keung \cite{bck} with a rather surprising result:  it has
a branching fraction of about $10^{-5}$.
This is almost two orders of magnitude larger than $Z^0 \rightarrow \psi g g$,
in spite of the fact that both rates are the same order in $\alpha_s$.
An explanation for the relatively large branching fraction
of $Z^0 \rightarrow \psi c {\bar c}$ was provided in Ref.
\cite{by}, where it was pointed out that this process includes a fragmentation
contribution that is not suppressed by a factor of $m_c^2/M_Z^2$.
This contribution can be factored into
the rate for the $Z^0$ to decay into a $c {\bar c}$ pair
multiplied by the probability for the $c$ or ${\bar c}$ to fragment
into $\psi$.

In this paper, we calculate the fragmentation functions
$D_{c \rightarrow \psi}(z,\mu)$ and $D_{c \rightarrow \eta_c}(z,\mu)$
for a charm quark to split into an S-wave charmonium state.
The fragmentation functions at the scale $\mu = m_c$ are calculated to leading
order in $\alpha_s(m_c)$.  Altarelli-Parisi equations are used to evolve
them up to the scale $\mu = M_Z/2$ appropriate for $Z^0$ decay.
Our simple analytic result for $Z^0 \rightarrow \psi c {\bar c}$
agrees with the complete leading order calculation of Barger, Cheung,
and Keung.  We also use our fragmentation functions
to calculate the direct production rates for $\psi$ in $W^\pm$ decays and
for $\Upsilon$ in top quark and Higgs decays.

\bigskip
{\bf \centerline{$Z^0$ Decay via Fragmentation}}

The fragmentation contribution to the inclusive decay rate of the $Z^0$
into charmonium is the term that survives in the limit
$M_Z/m_c \rightarrow \infty$.  The general form of the fragmentation
contribution to the differential decay rate for the production
of a $\psi$ of energy $E$ is
\begin{equation} {
d\Gamma ( Z^0 \rightarrow \psi(E) + X)
\;=\; \sum_i \int_0^1 dz \;
d{\widehat \Gamma}(Z^0 \rightarrow i(E/z) + X,\mu)
	\; D_{i \rightarrow \psi}(z,\mu) \;,
} \label{facZ} \end{equation}
where the sum is over partons of type $i$ and $z$ is the
longitudinal momentum fraction of the $\psi$ relative to the parton.
The physical interpretation of (\ref{facZ}) is that a $\psi$
of energy $E$ can be produced by first producing a parton $i$ of larger
energy $E/z$ which subsequently splits into a $\psi$ carrying
a fraction $z$ of the parton energy.
The expression (\ref{facZ}) for the differential decay rate has a factored
form:  all the dependence on the energy $E$ is in the parton subprocess
decay rate $d{\widehat \Gamma}$, while all the dependence on the charm
quark mass $m_c$ is in the fragmentation function $D_{i \rightarrow \psi}$.
To maintain this factored form in spite of the logarithms of $M_Z/m_c$
that arise in perturbation theory, a factorization scale $\mu$ must
be introduced.  The dependence on the arbitrary
scale $\mu$ cancels between the two factors.
Large logarithms of $E/\mu$ in the subprocess decay rate
${\widehat \Gamma}$ can be avoided by choosing $\mu$ on the order of $E$.
Large logarithms of $\mu/m_c$ then necessarily appear in the
fragmentation functions $D_{i \rightarrow \psi}(z,\mu)$,
but they can be summed up by solving the evolution equations \cite{rdf}
\begin{equation} {
\mu {\partial \ \over \partial \mu} D_{i \rightarrow \psi}(z,\mu)
\;=\; \sum_j \int_z^1 {dy \over y} \; P_{i\rightarrow j}(z/y,\mu)
	\; D_{j \rightarrow \psi}(y,\mu) \;,
} \label{evol} \end{equation}
where $P_{i\rightarrow j}(x,\mu)$ is the Altarelli-Parisi
function for the splitting of the parton of type $i$ into a parton of
type $j$ with longitudinal momentum fraction $x$.
For example, the $c \rightarrow c$ splitting function for
a charm quark with energy much greater than its mass is the usual
splitting function for quarks:
\begin{equation} {
P_{c \rightarrow c}(x,\mu) \;=\; {\alpha_s(\mu) \over 2 \pi}
	\left( {4 \over 3} \; {1 + x^2 \over (1-x)_+}
	\;+\; 2 \; \delta(1-x) \right) \;.
} \label{Pc} \end{equation}
The boundary condition on the evolution equation
(\ref{evol}) is the initial fragmentation function
$D_{i \rightarrow \psi}(z,m_c)$ at the scale $m_c$.  As shown in
Ref. \cite{by}, it can be calculated
perturbatively as a series in $\alpha_s(m_c)$.

We can easily count the order in $\alpha_s$ for the
fragmentation contributions to
$\psi$ production in $Z^0$ decay.  The subprocess rate ${\widehat \Gamma}$
for producing gluons is of order $\alpha_s$, while that for
producing quarks is of order 1.
The fragmentation function for a gluon to split into  $\psi$,
which was calculated in
Ref. \cite{by}, is proportional to $\alpha_s^3$.  A light quark
can split into a $\psi$ only by radiating a gluon which splits into a
$\psi$, so its fragmentation function is of order $\alpha_s^4$.  In contrast,
the fragmentation function for a charm quark to split into a $\psi$,
which will be calculated explicitly below, is only of order $\alpha_s^2$.
Thus the fragmentation of charm quarks into $\psi$ dominates by two powers
of $\alpha_s$ over the fragmentation of light quarks or gluons.

Keeping only the charm quark and antiquark contributions to
(\ref{facZ}), the energy distribution of the $\psi$ reduces
at leading order in $\alpha_s$ to
\begin{equation} {
{d\Gamma \over dz}( Z^0 \rightarrow \psi(E) + X)
\;=\;   2 \; {\widehat \Gamma}(Z^0 \rightarrow c {\bar c})
	\; D_{c \rightarrow \psi}\left(z,M_Z/2\right) \;,
\; z \;=\; {2E \over M_Z} \;.
} \label{facE} \end{equation}
This fragmentation formula is of course applicable only for a $\psi$ of
energy $E$ that is a significant fraction $z$ of the energy $M_Z/2$
of the charm quark and much greater than the mass $M_\psi$ of the $\psi$.
In (\ref{facE}), the factor of 2 accounts for
the contribution from the fragmentation of the ${\bar c}$.
We have set the factorization scale $\mu$ to
$M_Z/2$ to avoid large logarithms from higher orders in perturbation
theory.  At leading order in
$\alpha_s$, only the diagonal term in the evolution equation (\ref{evol})
survives:
\begin{equation} {
\mu {\partial \ \over \partial \mu} D_{c \rightarrow \psi}(z,\mu)
\;=\; \int_z^1 {dy \over y} \; P_{c\rightarrow c}(z/y,\mu)
	\; D_{c \rightarrow \psi}(y,\mu) \;.
} \label{evolc} \end{equation}
Integrating (\ref{facE}) over the energy, the total rate for
inclusive $\psi$-production is
\begin{equation} {
\Gamma( Z^0 \rightarrow \psi + X)
\;=\;  2 \; {\widehat \Gamma}(Z^0 \rightarrow c {\bar c})
	\; \int_0^1 dz \; D_{c \rightarrow \psi}(z,m_c) \;.
} \label{facpsi} \end{equation}
We have set the fragmentation scale equal to $m_c$
by exploiting the fact that at leading order in $\alpha_s$
the Altarelli-Parisi splitting function
(\ref{Pc}) satisfies $\int_0^1 dx P_{c \rightarrow c}(x,\mu) = 0$.
The evolution equation (\ref{evolc}) then implies that
the fragmentation probability
$\int_0^1 dz D_{c \rightarrow \psi}(z,\mu)$
does not evolve with the scale $\mu$.

\bigskip
{\bf \centerline{Fragmentation function for $c \rightarrow \psi$}}

We proceed to calculate the initial fragmentation function
$D_{c \rightarrow \psi}(z,m_c)$ for a charm quark to split into
a $\psi$ to leading order in $\alpha_s(m_c)$.
Our strategy is to isolate the contribution $\Gamma_1$ to the decay rate
for $Z^0 \rightarrow \psi c {\bar c}$ that arises from the fragmentation of
the charm quark.  We can then obtain the fragmentation probability
$\int_0^1 dz D(z)$ by
dividing $\Gamma_1$ by the rate $\Gamma_0$ for $Z^0 \rightarrow c {\bar c}$:
\begin{equation} {
\Gamma_0 \;=\; {1 \over 2 M_Z}
\int [d {\bar q}] [dq] \; (2 \pi)^4 \delta^4(Z - {\bar q} - q) \;
{1 \over 3} \; \sum |A_0|^2 \;,
} \label{Gzero} \end{equation}
where $Z$, ${\bar q}$, and $q$ are the 4-momenta of the $Z^0$, ${\bar c}$,
and $c$, and $[dq] = d^3q/(16 \pi^3 q_0)$
is the Lorentz-invariant phase space element.
The square of the amplitude $A_0$ for $Z^0 \rightarrow c {\bar c}$,
averaged over initial spins and summed over
final spins and colors, is
\begin{equation} {
{1 \over 3} \sum |A_0|^2 \;=\;
\left( - g^{\alpha \beta} + {Z^\alpha Z^\beta \over M_Z^2} \right)
{\rm tr} \Bigg( \Gamma_\alpha \; ({\not \! {\bar q}} - m_c) \;
		\Gamma_\beta \; ({\not \! q} + m_c) \Bigg) \;,
} \label{Azero} \end{equation}
where $\Gamma_\alpha$ is the $Z^0 c {\bar c}$ vertex whose explicit
form is not required.
In the limit $M_Z >> m_c$, the factors of $m_c$ in the trace can be neglected.

The rate for the decay $Z^0 \rightarrow \psi c {\bar c}$ is
\begin{equation} {
\Gamma_1 \;=\; {1 \over 2 M_Z}
\int [d{\bar q}] [dp] [dp'] \; (2 \pi)^4 \delta^4(Z - {\bar q} - p - p')
\; {1 \over 3} \sum |A_1|^2 \;,
} \label{Gone} \end{equation}
where ${\bar q}$, $p$, and $p'$ are the 4-momenta of the
${\bar c}$, $\psi$, and $c$.
The four Feynman diagrams that contribute to the amplitude $A_1$ at
leading order in $\alpha_s$ are shown in Figure 1.
The contributions to the process $Z^0 \rightarrow \psi c {\bar c}$
that correspond to the
fragmentation of the charm quark come from the region of phase space in
which the $\psi-c$ system has large momentum $q = p+p'$ of order $M_Z$
and small invariant mass $s = q^2$ of order $m_c^2$.
To facilitate the extraction of the fragmentation probability,
we write the 3-body phase space for the outgoing particles in
an iterated form by introducing integrals over $q$ and over $s$:
\begin{eqnarray}
&& \int [d{\bar q}] [dp] [dp'] \; (2 \pi)^4 \delta^4(Z - {\bar q} - p - p')
\nonumber \\
&=& \int {d s \over 2 \pi} \int [d {\bar q}] [dq]
	\; (2 \pi)^4 \delta^4(Z - {\bar q} - q)
	\int [dp] [dp'] \; (2 \pi)^4 \delta^4(q - p - p') \;.
\label{itphsp} \end{eqnarray}
We also express the two-body phase space integral over $p$ and $p'$
in terms of the longitudinal momentum fraction $z$ of the $\psi$.
In a frame in which the virtual charm quark has the
4-momentum $q =(q_0,0,0,q_3)$, the longitudinal momentum fraction of the $\psi$
is $z = (p_0 + p_3)/(q_0 + q_3)$ and its transverse momentum is
${\vec p}_\perp = (p_1, p_2)$.  Expressed in terms of these variables,
the Lorentz invariant phase space element is
$[dp] = dz d^2p_\perp /(16 \pi^3 z)$.  Integrating over
the 4-momentum $p'$ and
over ${\vec p}_\perp$, the 2-body phase space integral reduces to
\begin{equation} {
\int [dp] [dp'] \; (2 \pi)^4 \delta^4 (q - p - p')
\;=\; {1 \over 8 \pi} \int_0^1 dz
\; \theta \left( s - {4 m_c^2 \over z}
	- {m_c^2 \over 1-z} \right) \;.
} \label{phsptwo} \end{equation}
We have set $M_\psi = 2 m_c$, which is accurate
up to relativistic corrections.
If $s = q^2$ is of order $m_c^2$, the delta function $\delta^4(q-p-p')$
constrains ${\vec p}_\perp$ to be of order $m_c$.  From the mass-shell
condition, the component $p_0 - p_3 = (p_\perp^2 + 4 m_c^2)/(p_0+p_3)$
is of order $m_c^2/M_Z$.  Thus, to leading order in $m_c/M_Z$, we
can set $p = zq$.

We proceed to isolate the contribution to the amplitude $A_1$
{}from the fragmentation of the charm quark.
In covariant gauges, this contribution comes from both of the
diagrams in Figure 1a and 1b, while the diagrams in Figures 1c and 1d
contain contributions from ${\bar c}$ fragmentation.
In the axial gauge associated with the 4-vector
${\bar q}$, the contribution from fragmentation of the charm quark
comes only from the diagram shown in Figure 1a.
The amplitude for Figure 1a in this gauge can be reduced to
\begin{eqnarray}
A_1 \;=\;  {4 g_s^2 R(0) \over 3 \sqrt{6 \pi m_c}}
\epsilon_\alpha(Z) \epsilon_\mu(p)^* {1 \over (s - m_c^2)^2}
\; {\bar u}(p') \Bigg( 2m_c \; \gamma^\mu \; ({\not \! q} + m_c)
\nonumber \\
\;+\; {s - m_c^2 \over {\bar q} \cdot (2q - p)} \;
	{\not \! {\bar q}} \; \gamma^\mu \; ({\not \! p} + 2m_c)
\Bigg) \; \Gamma^\alpha \; v({\bar q}) \;.
\label{Apsi} \end{eqnarray}
We have used standard covariant Feynman rules \cite{gkpr} for projecting the
amplitude for production of a $c {\bar c}$ pair with equal 4-momenta
$p/2$ onto the amplitude for production of a $\psi$ with 4-momentum $p$.
The parameter $R(0)$ is the value of the nonrelativistic radial wavefunction
at the origin.
Averaging over inital spins and summing over final spins and colors,
the square of the amplitude reduces to
\begin{equation} {
{1 \over 3} \sum |A_1|^2 \;=\; {128 \pi \alpha_s^2 |R(0)|^2 \over 27 m_c} \;
{1 \over (s - m_c^2)^4}
\left( - g^{\alpha \beta} + {Z^\alpha Z^\beta \over M_Z^2} \right)
{\rm tr} \Bigg( \Gamma_\alpha \; ({\not \! {\bar q}} - m_c) \;
		\Gamma_\beta \; D \Bigg) \;,
} \label{Asq} \end{equation}
where $D$ is a Dirac matrix that depends on ${\bar q}$, $q$, and $p$.
We need only keep the terms in $D$ for which the Dirac trace in
(\ref{Asq}) is of order $m_c^4 M_Z^2$.
While ${\bar q}$, $q$, and $p$ all have components of order $M_Z$,
$s = q^2$ is of order $m_c^2$ in the fragmentation region.
Simplifying the Dirac matrix by dropping terms which are suppressed by
powers of $m_c/M_Z$, it reduces to
\begin{eqnarray}
D &=& (s^2 - 2 m_c^2s - 47 m_c^4) \; {\not \! q}
	\;-\; (s - m_c^2) (s - 9 m_c^2) \; {\not \! p}
\nonumber \\
&+& 4 \; {s-m_c^2 \over {\bar q} \cdot (2q-p)}
\left( (s+7m_c^2) \; {\bar q} \cdot q \; {\not \! p}
	\;-\; (s-5m_c^2) \; {\bar q} \cdot p \; {\not \! p}
	\;-\; 8 m_c^2 \; {\bar q} \cdot q \; {\not \! q} \right)
\nonumber \\
&+& 12 \left( {s-m_c^2 \over {\bar q} \cdot (2q-p)} \right)^2
	{\bar q} \cdot p \; {\bar q} \cdot (q-p) \; {\not \! p} \;.
\label{Dirpsi} \end{eqnarray}
We have exploited the fact that while ${\not \! p}$ and ${\not \! q}$
are both of order $M_Z$, their product ${\not \! p} {\not \! q}$
is only of order $m_c M_Z$.
The coefficients of ${\not \! p}$ and ${\not \! q}$
in (\ref{Dirpsi}) are all manifestly of order $m_c^4$, so we can
substitute $p = zq$ for all the remaining factors of $p$.
The Dirac trace in (\ref{Apsi}) is then proportional to
${\rm tr}(\Gamma_\alpha {\not \! {\bar q}} \;\Gamma_\beta {\not \! q})$.
It is now easy to divide $\Gamma_1$ by the decay rate $\Gamma_0$
given in (\ref{Gzero})  to obtain the fragmentation probability:
\begin{eqnarray}
\int_0^1 dz \; D_{c \rightarrow \psi}(z)
\;=\; {8 \alpha_s^2 |R(0)|^2 \over 27 \pi m_c}
\int_0^\infty ds \; {1 \over (s - m_c^2)^4} \int_0^1 dz \;
\theta \left(s - {4m_c^2 \over z} - {m_c^2 \over 1-z} \right)
\nonumber \\
\Bigg( (s^2 - 2 m_c^2 s - 47 m_c^4) \;-\; z (s - m_c^2) (s - 9 m_c^2)
\;+\; 4 {z(1-z) \over 2-z} s (s - m_c^2)
\nonumber \\
\;-\; 4 {8-7z-5z^2 \over 2-z}m_c^2 (s - m_c^2)
\;+\; 12 {z^2 (1-z) \over (2-z)^2} (s - m_c^2)^2 \Bigg) \;.
\label{Dint} \end{eqnarray}
Note that the upper limit on the integral over $s$
has been increased to $\infty$.
Since the integrand behaves like $1/s^2$ at large $s$, this only changes
the integral by an amount of order $m_c^2/M_Z^2$, which
we have been systematically neglecting.
Evaluating the integral over $s$ in (\ref{Dint}),
we obtain our final expression
for the initial fragmentation function:
\begin{equation}{
D_{c \rightarrow \psi}(z, 3 m_c)
\;=\; {64 \over 27 \pi} \; \alpha_s(3 m_c)^2  \;
{|R(0)|^2 \over M_\psi^3} \;
{z (1-z)^2 (16 - 32 z + 72 z^2 - 32 z^3 + 5 z^4)\over (2-z)^6} \;.
} \label{Dpsi} \end{equation}
We have set the scale $\mu$ in the fragmentation function
and in the running coupling constant to $\mu = 3 m_c$, which is the
minimum value of the invariant mass $\sqrt{s}$ of the fragmenting charm quark.
We have also set $2 m_c \rightarrow M_\psi$ in the denominator,
which is accurate up to relativistic corrections.
Integrating over $z$, we obtain the total fragmentation probability:
\begin{equation} {
\int_0^1 dz \; D_{c \rightarrow \psi}(z, 3 m_c) \;=\;
{64 \over 27 \pi} \; \alpha_s(3 m_c)^2  \; {|R(0)|^2 \over M_\psi^3} \;
\left( {1189 \over 30} - 57 \log 2 \right) \;.
} \label{Ppsi} \end{equation}

\bigskip
{\bf \centerline{Fragmentation function for $c \rightarrow \eta_c$}}

The fragmentation function for a charm quark to split into the $^1S_0$
state of charmonium $\eta_c$ can be calculated in the same way as for $\psi$.
The starting point is the expression (\ref{Gone})
for the decay rate for $Z^0 \rightarrow \eta_c c {\bar c}$, except that the
amplitude $A_1$ in (\ref{Apsi}) must be replaced by
\begin{eqnarray}
A_1 \;=\; {4 g_s^2 R(0) \over 3 \sqrt{6 \pi m_c}}
\epsilon_\alpha(Z) \epsilon_\mu(p)^* {1 \over (s - m_c^2)^2}
{\bar u}(p') \Bigg( ({\not \! p}  + 4 m_c) \; \gamma_5 \;
	({\not \! q} + m_c)
\nonumber \\
\;+\; {s - m_c^2 \over {\bar q} \cdot (2q - p)} \; {\not \! {\bar q}} \;
	\gamma_5 \; ({\not \! p} + 2m_c)
\Bigg) \Gamma^\alpha v({\bar q}) \;.
\label{Aeta} \end{eqnarray}
The square of the amplitude has the form (\ref{Asq}), except that the
Dirac matrix $D$ reduces to
\begin{eqnarray}
D &=& (s + 3 m_c^2)(s - 5 m_c^2) \; {\not \! q}
	\;-\;
(s - m_c^2) (s - 9 m_c^2) \; {\not \! p}
\nonumber \\
&+& 4 {s-m_c^2 \over {\bar q} \cdot (2q-p)}
\left( (s-m_c^2) \; {\bar q} \cdot q \;
	\;-\; (s - 3m_c^2) \; {\bar q} \cdot p \; \right) {\not \! p}
\nonumber \\
&+& 4 \left( {s-m_c^2 \over {\bar q} \cdot (2q-p)} \right)^2
	{\bar q} \cdot p \; {\bar q} \cdot (q-p) \; {\not \! p} \;.
\label{Direta} \end{eqnarray}
Following the same path as in the $\psi$ calculation, we find that
the initial fragmentation function for $\eta_c$ is
\begin{equation} {
D_{c \rightarrow \eta_c}(z, 3 m_c)
\;=\; {64 \over 81 \pi} \; \alpha_s(3 m_c)^2  \;
{|R(0)|^2 \over M_{\eta_c}^3} \;
{z (1-z)^2 ( 48 + 8 z^2 - 8 z^3 + 3 z^4 )\over (2-z)^6} \;.
} \label{Deta} \end{equation}
Integrating over $z$, the fragmentation probability is
\begin{equation} {
\int_0^1 dz \; D_{c \rightarrow \eta_c}(z, 3 m_c) \;=\;
{64 \over 27 \pi} \; \alpha_s(3 m_c)^2  \; {|R(0)|^2 \over M_{\eta_c}^3} \;
\left( {773 \over 30} \;-\; 37 \log 2 \right) \;.
} \label{Peta} \end{equation}

\bigskip
{\bf \centerline{Decay of $Z^0$ into Charmonium}}

{}From (\ref{facpsi}), the branching ratio for the decay of the $Z^0$
into $\psi$ relative to the decay into $c {\bar c}$ is
\begin{equation} {
{\Gamma(Z^0 \rightarrow \psi c {\bar c}) \over
	\Gamma(Z^0 \rightarrow c {\bar c})}
\;=\; 0.1870 \; \alpha_s(3 m_c)^2  \; {|R(0)|^2 \over M_\psi^3} \;
} \label{BRpsi} \end{equation}
The value of the parameter $R(0)$ can be determined from the $\psi$
electronic width to be $|R(0)|^2 = (0.82 \; {\rm GeV})^3$.
Taking $\alpha_s(3 m_c) = 0.23$, we find that the branching ratio
(\ref{BRpsi}) is $1.8 \times 10^{-4}$.  The simple result (\ref{BRpsi})
agrees with the complete leading order calculation of
$Z^0 \rightarrow \psi c {\bar c}$ in Ref. \cite{bck} after taking into
account the differences in the values of $R(0)$, $\alpha_s$,
and the charm quark mass.  A slightly larger
value for the wavefunction at the origin was used in Ref. \cite{bck}:
$|R(0)|^2 = (0.92 \; {\rm GeV})^3$.  It was also assumed
implicitly in Ref. \cite{bck} that $Z^0 \rightarrow \psi c {\bar c}$
is a short distance process, so the running coupling constant was
taken to be $\alpha_s(M_Z) \approx 0.15$.  As we have shown, the
dominant contribution comes from a fragmentation process, and the
appropriate scale of the coupling constant is definitely on the order of $m_c$.
Finally, instead of $M_\psi^3$ in the denominator of (\ref{BRpsi}),
the authors of Ref. \cite{bck}
used $(2 m_c)^3$ with $m_c = 1.35 \; {\rm GeV}$.
The difference between $M_\psi$ and $2 m_c$ is a relativistic correction,
which we have consistently ignored in this analysis.
Corrections to the fragmentation approximation are of order
$(2 M_\psi/M_Z)^2$ or about 0.4\%, which is much smaller than the size
of relativistic corrections and higher order perturbative corrections.

The rate for production of $\eta_c$ by fragmentation differs by less than 3\%
{}from that for $\psi$.  From (\ref{Peta}), we obtain
\begin{equation} {
{\Gamma(Z^0 \rightarrow \eta_c c {\bar c}) \over
	\Gamma(Z^0 \rightarrow c {\bar c})} \;=\;
0.1814 \; \alpha_s(3 m_c)^2  \; {|R(0)|^2 \over M_{\eta_c}^3} \;
} \label{BReta} \end{equation}
This agrees with the calculation of Ref. \cite{bck}
after taking into account the differences in the values of
$\alpha_s$, $R(0)$, and $m_c$ and an apparent algebraic error
of a factor of 3.

The energy distribution of the $\psi$'s produced
by the fragmentation of charm quarks in $Z^0$ decay
is given in (\ref{facE}).  It is proportional to the fragmentation function
evaluated at the scale $M_Z/2$.  The initial fragmentation function
(\ref{Dpsi}) at the scale $3 m_c$ is shown as a solid line in Figure 1.
It must be evolved up to the scale $M_Z/2$ using the
Altarelli-Parisi equation (\ref{evolc}) in order to sum up the
leading logarithms of $M_Z/m_c$ from higher order radiative corrections.
The result is shown as the dotted line in Figure 2.
The evolution softens the energy distribution,
shifting the peak in the fragmentation function from
$z = 0.75$ to  $z = 0.68$.
The energy distribution shown in Figure 2 should be accurate provided that the
energy $E$ of the $\psi$ is large compared to its mass, or
equivalently $z >> 0.07$.
The fragmentation function for $\eta_c$
production is also shown in Figure 2.  It has a slightly softer
distribution, but its behavior is otherwise similar to that for the $\psi$.

The expression (\ref{BRpsi}) also applies with minor modifications to
the corresponding branching ratio for $\Upsilon$ production:
\begin{equation} {
{\Gamma(Z^0 \rightarrow \Upsilon b {\bar b}) \over
	\Gamma(Z^0 \rightarrow b {\bar b})} \;=\;
0.1870 \; \alpha_s(3 m_b)^2  \; {|R(0)|^2 \over M_\Upsilon^3} \;
} \label{BRUps} \end{equation}
where $R(0)$ is the radial wavefunction at the origin for the $\Upsilon$,
which is determined from its electronic decay rate
to be $|R(0)|^2 = (1.72 \; {\rm GeV})^3$.
Taking $\alpha_s(3 m_b) =0.17$, we find that the branching ratio
(\ref{BRUps}) is $3.3 \times 10^{-5}$. The fragmentation approximation for
$\Upsilon$ production in $Z^0$ decay is not as accurate as it is for
$\psi$ production.  Corrections are on the order of
$(2 M_\Upsilon/M_Z)^2$, which is about 4\%.

\bigskip
{\bf \centerline{Decay of $W^\pm$ into $\psi$}}

About $1/3$ of the decays of the $W^+$ will proceed through the channel
$W^+ \rightarrow c {\bar s}$. The mass of the $W$ is sufficiently large
that the dominant production mechanism for charmonium will be
$W^+ \rightarrow c {\bar s}$, followed by the fragmentation of the
charm quark into charmonium.  Fragmentation of the strange antiquark into
$\psi$ is suppressed by a factor of $\alpha_s^2$.
The branching ratio for decay into $\psi$ relative to decay into $c {\bar s}$
is therefore smaller than (\ref{BRpsi}) by a factor of 2:
\begin{equation} {
{\Gamma(W^+ \rightarrow \psi c {\bar s}) \over
	\Gamma(W^+ \rightarrow c {\bar s})} \;=\;
0.0935 \; \alpha_s(3 m_c)^2  \; {|R(0)|^2 \over M_\psi^3} \;.
} \label{BRW} \end{equation}
Numerically this branching ratio is $9.2 \times 10 ^{-5}$.
Our analytic calculation of the fragmentation contribution
is consistent with the full leading order calculation of Ref. \cite{bck}.

\bigskip
{\bf \centerline{Decay of Top Quark into $\Upsilon$}}

The top quark will probably decay almost exclusively into $W^+ b$.
If the top quark is heavy enough, the dominant production mechanism
for bottomonium in top quark decay will be $t \rightarrow W^+ b$,
followed by the fragmentation of the $b$ quark into bottomonium.
The branching fraction for the direct decay into the $^3S_1$ state
$\Upsilon$ is one half of (\ref{BRUps}):
\begin{equation} {
{\Gamma(t\rightarrow W^+ \Upsilon b) \over \Gamma(t \rightarrow W^+ b)}
\;=\;
0.0935 \; \alpha_s(3 m_b)^2  \; {|R(0)|^2 \over M_\Upsilon^3} \;,
} \label{BRtop} \end{equation}
which has the numerical value $1.6 \times 10^{-5}$. The
complete leading order calculation of the rate
for $t \rightarrow W^+ b \Upsilon$ gives a branching fraction of
$4 \times 10^{-7}$ for a top quark with a mass of 100 GeV \cite{bck}.
The fragmentation formula (\ref{BRtop}) does not apply to such a small
value of the top quark mass, since the maximum momentum of the
$\Upsilon-b$ system is only 13 GeV, too small
for the decay rate to be dominated by fragmentation.  The simple
result (\ref{BRtop}) is a good approximation if the mass
of the top quark is closer to 150 GeV.

\bigskip
{\bf \centerline{Decay of Higgs into $\Upsilon$}}

If the Higgs mass is  below the threshold for decay into
$W$ pairs, than its dominant decay mode will be $H \rightarrow b {\bar b}$.
The dominant production method
for bottomonium in Higgs decay will be $H \rightarrow b {\bar b}$,
followed by the fragmentation of the $b$ quark or antiquark into bottomonium.
The branching fraction for the direct decay into the
$\Upsilon$ is twice (\ref{BRtop}), because both the $b$ and ${\bar b}$
can fragment into $\Upsilon$:
\begin{equation} {
{\Gamma(H \rightarrow \Upsilon b {\bar b}) \over
	\Gamma(H \rightarrow b {\bar b})} \;=\;
0.1870 \; \alpha_s(3 m_b)^2  \; {|R(0)|^2 \over M_\Upsilon^3} \;,
} \label{BRH} \end{equation}
This branching ratio is $3.3 \times 10^{-5}$,
which is probably too small
for this decay mode to be useful as a signal
for an intermediate mass Higgs boson.

\bigskip
{\bf \centerline{Conclusions}}

We have shown in this paper that the dominant mechanism
for the direct production of charmonium in $Z^0$ decay is fragmentation,
the production of a high energy charm quark
or antiquark followed by its splitting into
the charmonium state.  Most previous calculations of
charmonium production have considered only short-distance production
mechanisms which are suppressed by a factor of $m_c^2/M_Z^2$.
We calculated the fragmentation functions
$D(z,\mu)$ for charm quarks or antiquarks to split into S-wave
charmonium states to leading
order in $\alpha_s$.  The fragmentation functions satisfy
Altarelli-Parisi evolution equations which can be used to sum up
large logarithms of $M_Z/m_c$.  These fragmentation functions are universal,
applying to the production of heavy quarkonium in any high energy
process that can produce heavy quarks with energy large compared to their
mass.  We applied them to the production of charmonium and bottomonium
in decays of the $Z^0$, $W^\pm$, top quark, and Higgs boson.

A complete calculation of the rate for $\psi$
production in $Z^0$ decay must include the production of the P-wave
charmonium states $\chi_{cJ}$, followed by their radiative decays into $\psi$.
In calculating the production of
P-wave charmonium states, there are two distinct contributions that must
be included at leading order in $\alpha_s$.
The P-wave state can arise either from the
production  of a collinear $c {\bar c}$ pair in a color-singlet P-wave state,
or from the production of a collinear $c {\bar c}$ pair in a color-octet
S-wave state \cite{bbly}.  The calculation of the fragmentation functions
for the splitting of charm quarks into P-wave charmonium states
will be presented elsewhere \cite{bya}.

This work was supported in part by the U.S. Department of Energy,
Division of High Energy Physics, under Grant DE-FG02-91-ER40684.
\vfill\eject

\noindent{\Large\bf Figure Captions}
\begin{enumerate}
\item The four Feynman diagrams for $Z^0 \rightarrow \psi c {\bar c}$
	at leading order in $\alpha_s$.
\item The fragmentation functions $D_{c \rightarrow \psi}(z,\mu)$
	and $D_{c \rightarrow \eta_c}(z,\mu)$
	as a function of $z$ for $\mu = 3 m_c$ (solid lines)
	and $\mu = M_Z/2$ (dotted lines).
\end{enumerate}
\vfill\eject


\begin{thebibliography}{99}
%
\bibitem{gkpr}
{B. Guberina, J.H. K\"uhn, R.D. Peccei, and R. R\"uckl,
		{\it Nucl. Phys.} {\bf B174}, 317 (1980).}
%
\bibitem{keung}
{W.-Y. Keung, {\it Phys. Rev.} {\bf D23}, 2072 (1981).}
%
\bibitem{abr}
{K.J. Abraham, {\it Z. Phys..} {\bf C44}, 467 (1989).}
%
\bibitem{ks}
{J.H. K\"uhn and H. Schneider, {\it Phys. Rev.} {\bf D24}, 2996 (1981);
	{\it Z. Phys.} {\bf C11}, 263 (1981).}
%
\bibitem{by}{E. Braaten and T.C. Yuan,
	Northwestern preprint NUHEP-TH-92-23 (November 1992).}
%
\bibitem{hms}{K. Hagiwara, A.D. Martin, and W.J. Stirling,
		{\it Phys. Lett.} {\bf B267}, 527 (1991)
		and erratum (to be published).}
%
\bibitem{bck}{V. Barger, K. Cheung, and W.-Y. Keung,
		{\it Phys. Rev.} {\bf D41}, 1541 (1990).}
%
\bibitem{rdf}{R.D. Field, {\it Applications of Perturbative QCD}
		(Addison Wesley, 1989).}
%
\bibitem{bbly}{G.T. Bodwin, E. Braaten, T.C. Yuan, and G.P. Lepage,
		{\it Phys. Rev.} {\bf D46}, R3703 (1992).}
%
\bibitem{bya}{E. Braaten, K. Cheung, and T.C. Yuan (in preparation).}
%
\end{thebibliography}
\end{document}